**Graphic abstract**

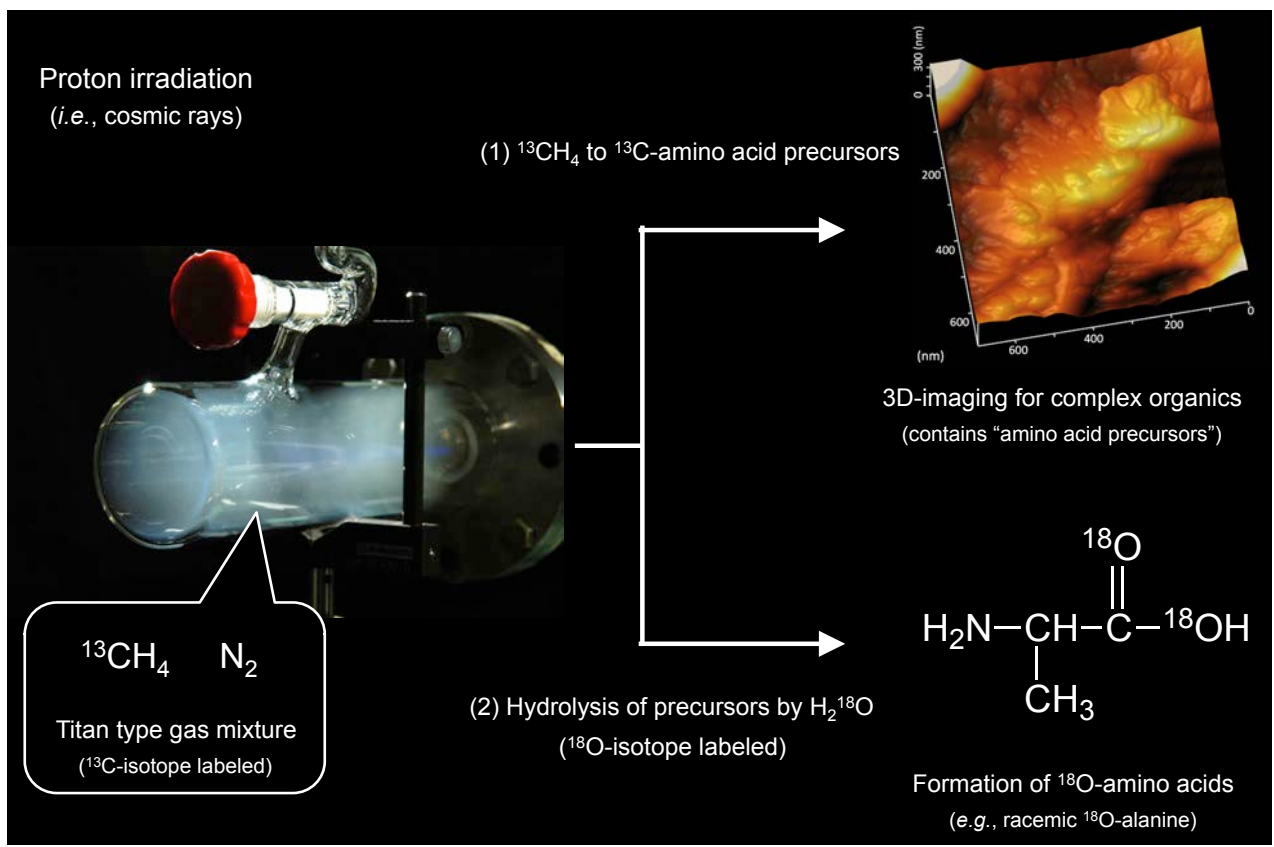

**Amino Acid Precursors from a Simulated Lower Atmosphere of Titan: Experiments of Cosmic Ray Energy Source with $^{13}$C- and $^{18}$O- Stable Isotope Probing Mass Spectrometry**

5   **Toshinori Taniuchi*, Yoshinori Takano**, and Kensei Kobayashi* †**

*Department of Chemistry and Biotechnology,

Yokohama National University,

79-5 Tokiwadai, Hodogaya-ku, Yokohama 240-8501, Japan

** Institute of Biogeosciences,

Japan Agency for Marine-Earth Science and Technology (JAMSTEC),

2-15 Natsushima-cho, Yokosuka 237-0061, Japan

† To whom correspondence should be addressed.
E-mail: kkensei@ynu.ac.jp

**Abstract**

The organic haze of aerosols that shrouds the Saturnian moon Titan has previously been studied by both observations and laboratory simulation experiments. Here we report the abiotic formation of amino acid precursors in complex organic molecules during experimental simulation of the environment near Titan's surface with proton irradiation. Pyrolysis of the organic molecules formed in the simulated Titan atmosphere by proton irradiation at 600°C yielded compounds that contained HCN and $NH_3$ ($m/z$ = 27 and 17). These experimental results are consistent with the molecular information obtained by pyrolysis gas chromatography/mass spectrometry (pyrolysis GC/MS) of samples collected by the Huygens probe to Titan. Scanning electron microscopy (SEM) and three-dimensional atomic force microscopy (AFM) images of the irradiation products reveal nanometer-scale filaments and globules in complex amorphous structures (approximately 1000 Da). Isotope probing experiments by matrix-assisted laser desorption ionization–time of flight–mass spectrometry (MALDI–TOF–MS) show that oxygen atoms were incorporated into the racemic amino acids by hydrolysis of $^{18}O$-labeled water. We suggest that the amino acid precursors possibly formed after water hydrolysis, as suggested in a previous observational study (C.A. Griffith, T. Owen, T.R. Geballe, J. Rayner, and P. Rannou, *Science*, **2003**, *300*, 628). We propose that cosmic rays are a significant and effective energy source for producing complex organics and amino acid precursors in Titan's atmospheric haze.

**Introduction**

50  Titan is the largest satellite of Saturn and has a dense atmosphere comprising mainly nitrogen ($N_2$) and methane ($CH_4$), as detected by ground-based and fly-by observational measurements in recent decades.[1] Initially, the Voyager spacecraft encounter in 1980 revealed the presence of an organic haze in the Titan's atmosphere that presumably included many organic molecules. The haze appeared to be caused by ultraviolet light and charged particles.[2–4]

55  Although Titan's surface is shrouded in a thick reddish haze,[5–8] direct and indirect observations suggested that liquid hydrocarbons and water (ice) were present at its surface.[9–13] The Huygens probe descended to Titan's surface from the Cassini spacecraft[14,15] in 2005 and transmitted data gathered from Titan's organic aerosol back to Earth.[16] These data included information about zonal winds,[17] along with the first views of Titan's surface and its spatial

60  variability.[18–20] Mass spectra obtained after pyrolysis of captured haze particles showed that the organic haze included –CN and –$NH_2$ functional groups.[16] The Cassini mission also identified a lake of liquid ethane and methane on the surface of Titan.[21]

Given that Titan resembles a primitive Earth with a dense nitrogen-rich atmosphere, numerous observations and simulation experiments have been performed to model the

65  processes that may have been involved in the chemical evolution of the origin of life, which was strongly dependent on the presence of UV light, electrical discharge, and heat as energy sources.[22–29] Organic materials produced in laboratory simulations of Titan's atmosphere are called "tholins", and include hydrocarbons, nitriles, heterocyclic aromatic compounds, and other amino acid precursor molecules. Cosmic rays may have been potentially effective
70  energy sources for initial ionization and abiotic formation of organic compounds near Titan's surface (Figs. 1a and 1b).[26,30–36] However, few laboratory simulations have studied the abiotic formation of organic compounds by cosmic rays with the focus on amino acid precursors and its hydrolysis process.

In this study, we irradiated a mixture of nitrogen and methane using a 3 MeV proton
75  beam to simulate the possible formation of organic compounds in the lower Titan atmosphere (about 0–200 km) by penetrative particle energy sources (*i.e.*, cosmic rays; Fig.1). The produced tholins were first analyzed by online pyrolysis gas chromatography combined with mass spectrometry (pyrolysis GC/MS) to examine their chemical structure. Gel permeation chromatography (GPC) was then used to estimate the molecular weights of the products.
80  After hydrolysis, the amino acids produced were identified and quantified by high-performance liquid chromatography (HPLC), gas chromatography mass spectrometry (GC/MS), and matrix-assisted laser desorption ionization–time of flight–mass spectrometry (MALDI–TOF–MS), including $^{13}$C and $^{18}$O stable isotope probing. This result showed that oxygen atoms of amino acids were incorporated during hydrolysis of amino acid precursors in
85  the tholins. To our knowledge, this is the first report to validate abiotic synthesis of amino acids after hydrolysis of the precursors in simulated Titan tholin experiments.

**Experimental**

*Materials*

90          All compounds used in the experiments were analytical grade reagents. $^{18}$O-labeled water (H$_2$$^{18}$O) was purchased from Cambridge Isotope Laboratories, Inc. (USA). Ultrapure grade nitrogen and a mixture of nitrogen (N$_2$) and $^{12}$C-methane ($^{12}$CH$_4$) were obtained from Taiyo Nippon Sanso (Japan). $^{13}$C-labeled methane ($^{13}$CH$_4$) was purchased from Isotec (USA). The D- and L-enantiomers of the amino acids ethyl chloroformate (ECF), chloroform, and
95     sodium chloride were acquired from Wako Pure Chemical Industries (Japan). 2,2,3,3,4,4,4-heptafluoro-1-butanol (HFB) was purchased from Tokyo Chemical Industry (Japan). All water used in the experiments was deionized and further purified using a Millipore Milli-Q LaboSystem and a Millipore Simpli Lab-UV system (Japan Millipore, Tokyo, Japan) to remove inorganic ions and organic contaminants. All glassware was heated
100    in a high-temperature oven (Yamato DR-22, Japan) at 500°C for 4 hours to eliminate contamination from organic compounds.

*Proton irradiation experiment*

We prepared gas mixtures typical of the Titan atmosphere[30–32] (at 700 Torr) from
105    nitrogen (95%) and $^{12}$C-methane (5%) or $^{13}$C-labeled methane (5%) in Pyrex glass tubes (400 mL) with Havar foil windows. It should be noted that the pressure of the experimental gas mixture normalized as 700 Torr is 93100 Pa, *i.e.*, in the range of $10^4$ to $10^5$ Pa as shown in Fig.1a. The experimental condition for proton irradiation, therefore, is appropriate to simulate the interaction between the lower atmosphere of Titan and cosmic rays energy source. The gas
110    mixtures were irradiated at 298 K with high-energy protons (3 MeV; total energy deposited, 2.7 x $10^{22}$ eV per run[30]) emitted by a Van de Graff accelerator (Fig. 2) at the Tokyo Institute of Technology (TIT)[37–39] or a tandem accelerator at the Japan Atomic Energy Agency (JAEA).[40] The temperature conditions (298 K) were constrained by the experimental limitations of the accelerator facilities. After proton irradiation, the yellowish solid products were dissolved in 5 mL of tetrahydrofuran (THF) or 5 mL of water and recovered. Solubility of hydrophilic compounds in THF is very similar to that in water. Given that we simulated the hydrolysis of amino acid precursors to form amino acids, we did not use water during sample recovery prior to chromatographic procedures.

The solid products containing complex organic molecules were recovered and gently dried at ambient temperature and pressure for scanning electron microscopy (SEM; JEOL JSM-6700F) and atomic force microscopy (AFM; Seico Instruments, SII SPA 400 unit, Japan).[41,42] The microscopic imaging samples were recovered by water to observe the morphology of Tholin after eventual water exposure, as suggested by Griffith and co-workers.[10] We used these imaging techniques to observe the morphology of simulated Titan tholin products and to determine the three-dimensional size (10 nm to 10 μm) of the amorphous organic aggregates.

*Pyrolysis gas chromatography/mass spectrometry*

Aliquots of the tholins recovered in the THF fraction were analyzed by pyrolysis GC/MS (JEOL JMS-600, Frontier Lab Double-Shot Pyrolyzer PY-2020iD, column DB-5 from J & W Scientific, 30 m × 0.25 mm internal diameter, 0.25 μm film thickness). Pyrolysis was conducted at temperatures of 600°C or 250°C over run times of 1 min and the sample inlet temperature was 320°C. A splitless injection was used with a purge time delay of 0.5 min. The column temperature was 30°C and programmed to heat up at a rate of 1°C/min up to 40°C, 2°C/min up to 60°C, 3°C/min up to 100°C, 5°C/min up to 200°C, and 20°C/min up to 350°C. Helium was used as a carrier gas with a flow rate of 1.0 mL/min. Electron ionization (EI) at 70 eV was applied. The pyrolysis GC/MS was calibrated over an *m/z* range of 10 - 130.

*Gel permeation and ion exchange chromatography*

GPC analysis was used to estimate the molecular weight distributions of tholins recovered from the THF fraction (column TOSOH G2000 HXL 7.8 mm × 30 cm, carrier THF, wavelength 254 nm). The sample was divided into five aliquots according to retention time and molecular weight as calibrated by polystyrene standard reagents. The aliquots were hydrolyzed to quantify each amino acid after GPC separation by hydrolyzing each fraction using 6 M HCl at 110°C for 24 h in order to evaluate the yields. Amino acids were identified and quantified by ion exchange high-performance liquid chromatography (HPLC; Shimazu LC-10A series).[38–40]

*Enantiomeric analysis of the amino acids to obtain D/L ratios*

The $^{13}$C-stable isotope probing signatures from starting $^{13}$C-carbon source (*i.e.*, $^{13}$C-CH$_4$) to $^{13}$C-amino acid products (*e.g.*, $^{13}$C-glycine) were traced by the GC/MS analysis. Prior to gas chromatography, the hydrolyzed amino acid fraction was derivatized as follows[43,44]: (1) 100 μL of the amino acid fraction was transferred to a Mini-Vial (1 mL volume, GL Sciences, Japan); (2) 50 μL of a HFB–pyridine mixture (3:1, v/v) was added; (3)

after briefly agitating, 10 μL of ECF was added and the solution was mixed for 10 s; (4) 30 mL of chloroform and 10 mg NaCl were then added and the solution was again gently mixed for 10 s; (5) approximately 1.0 μL of the organic phase was then injected into the GC instrument. Amino acid enantiomers were identified by GC/MS (JEOL JMS-600, column Alltech Chirasil-L-Val 25 m × 0.25 mm internal diameter, 0.16 μm film thickness). The sample inlet temperature was 250°C. Splitless injection was used with a purge time delay of 0.7 min. The column temperature was 80°C, with a 1 min hold time, and the temperature was then programmed to increase at a rate of 5°C/min up to 150°C, followed by a 7 min hold time, and then another increase at a rate of 7°C/min up to 200°C. Helium was used as a carrier gas at a flow rate of 0.7 mL/min. Electron impact (EI) at 70 eV was used for detection.

*MALDI–TOF–MS analysis to determine stable isotope labeling*

The $^{18}$O-stable isotope probing signatures after $H_2{}^{18}O$ hydrolysis of amino acid precursors were assigned by MALDI–TOF–MS analysis (Shimadzu/Kratos Model AXMA-CFR).[31,32] MALDI–TOF–MS analysis of the samples was performed to identify the oxygen atoms from $^{18}$O-labeled water ($H_2{}^{18}O$) introduced to carboxylic and hydroxyl groups in the amino acids. The irradiation products were recovered in 5 mL $H_2{}^{18}O$. A 1.1 mL volume of 12 M HCl was added to a 0.1 mL aliquot of the sample, and followed by hydrolysis in 1 M HCl at 110°C for 24 h. Approximately 20 mg of α-cyano-4-hydroxycinnamic acid (CHCA) was used as the matrix for the TOF–MS analysis. TOF analysis was conducted in reflection mode and positive ion detection.

**Results and Discussion**

*Production of tholins and morphology of the pyrolysis products*

180    A white–yellow misty haze formed immediately (approximately 10 seconds) after starting proton irradiation (Figs. 2a and ab). Recovery of the solid products by rinsing with $H_2O$ or THF produced a yellow solution. Aliquots of the tholins were examined by SEM and AFM to constrain the morphology (Fig. 3) and size distribution (Fig. 4) of the solid products. SEM imaging showed that the tholins have filament and/or globular structures and an

185    amorphous component. The amorphous features (Figs. 3a and 3b) have irregular depressions, as evident in Figs. 4a and 4b. Three-dimensional AFM images of the tholin products reveal irregular surfaces with ragged hollows (>10 nm). Figures 3d, 3e and 4c, 4d are suggestive of tholin growth as aggregated amorphous soft matter. The thickness of the aggregates was estimated to be >100 nm by three-dimensional AFM imaging. In a cold dry environment like

190    that present on the surface of Titan, these complex organic molecules may condense and aggregate as aerosols.

   After online pyrolysis of the tholins at 600°C, peaks at *m/z* = 17 and 27 were the main signals detected in the mass spectrum (Fig. 5), implying the presence of $NH_3$ and HCN, respectively. These results indicate that $–NH_2$ and –CN chemical bonds are present in the

195    aerosol particles, which is consistent with the results of the Huygens Aerosol Collector and Pyrolyzer (ACP).[16] Similar mass spectra were obtained from samples heated at 250°C. The intensity of the main peak associated with the pyrolysis products at 250°C was 50% of the intensity of the peak observed at 600°C. The data obtained for heating at 250°C indicate that $NH_3$ and HCN had condensed onto the complex organic structures formed by the proton beam.

200 We conclude that tholins formed by proton irradiation include both bonding states noted above, and that the Huygens ACP could not detect the accumulated or adherent $NH_3$ or HCN molecules due to the small sample volume analyzed by Huygens.

Our preliminary report on a Curie-point pyrolysis GC/MS study of proton irradiated tholin products[30] showed that cyclic and heterocyclic compounds were also detected
205 (Supplementary Information), including alkylpyrroles (R–$C_4H_4N$), alkylpyridines (R–$C_5H_5N$), alkylpyradines (R–$C_4H_3N_2$), methylpyrimidine ($CH_3$–$C_4H_3N_2$), naphthalene, phenanthrene, anthracene, and quinoline/isoquinoline ($C_9H_7N$).

*Formation of amino acids after acid hydrolysis*

210 Thirteen amino acids including enantiomers were identified in the tholin samples after acid hydrolysis. Glycine was predominant with a G-value (number of molecules per 100 eV of deposited energy)[30, 38, 40] of 0.03, produced from an initial 5% methane gas sample and recovered using THF. GC/MS was used to determine the source of the amino acids (Fig. 6a). D- and L-enantiomers of the amino acids (alanine, α-aminobutyric acid, valine, and
215 norvaline) were detected as racemic components. The mass shift was measured using $^{13}CH_4$ as a carbon source (Fig. 6b). A peak at *m/z* = 329 corresponding to a molecular ion from derivatized glycine was shifted to *m/z* = 331 due to the incorporation of two $^{13}$C-atoms, suggesting $^{13}$C-glycine formation from $^{13}$C-$CH_4$. Figure 7 shows the MALDI–TOF mass spectra of glycine and alanine formed after acid hydrolysis of the tholins. When the tholins
220 were hydrolyzed in the presence of $H_2^{18}O$, a mass shift corresponding to the incorporation of two oxygen atoms into the carboxyl groups of glycine (*m/z* = 76) and alanine (*m/z* = 90) occurred and shifted the peaks to *m/z* = 80 and 94, respectively. These results show that oxygen atoms were incorporated during the production of amino acids by hydrolysis of the tholins.

225    The GPC results show that the tholins had a molecular weight up to 1000 Da (Fig. 8) as determined by comparison with polystyrene calibration standards. The retention time of the main peak indicates that a large fraction of the products have molecular weights ranging from 200 to 300 Da. Hence, a $^{13}$C-isotope probe was incorporated into a large fraction of the amino acids (Fig. 6b). These results imply that cosmic rays comprising high-energy protons
230    do not produce simple molecules, but complex organic compounds with molecular weights of several hundred Da.

**Conclusions**

(1) Using $^{13}$C and $^{18}$O stable isotope probing experiments, we have shown that abiotic
235    formation of $^{13}$C-amino acids from $^{13}$C-CH$_4$ and $^{18}$O-oxygen atoms of amino acids can be incorporated during hydrolysis of amino acid precursors produced by cosmic ray irradiation. Our results suggested that abiotic synthesis of amino acids can be produced after hydrolysis of amorphous amino acid precursors from simulated tholins with molecular weights of <1000 Da.
240    (2) Pyrolysis mass spectra for compounds formed under conditions that simulate those on the surface of Titan's moon are consistent with those obtained by the Huygens ACP, which suggests that cosmic rays provide an effective energy source for haze production in Titan's lower atmosphere. Tholins produced in the presence of cosmic rays could yield amino acids upon interaction with surface water/ice[10,12] or with cometary water/ice during a meteorite or

245        comet impact.[45–47]

      Given that rain from Titan's methane and ethane clouds could produce rivers and lakes, complex organic molecules may have potentially accumulated in the liquid methane flows on Titan's surface.[12,13] Furthermore, the presence of subsurface liquid water and ammonia,[48] methane drizzle,[49] along with various abiotic radiation products[50] is likely to

250      represent the next step in the chemical evolution on Titan. Possible future *in situ* investigations by Titan Mare Explorer (TiME)[51] hold considerable promise to further understand the pristine organic carbon and nitrogen cycles in the hydrology of Titan.

**Acknowledgements**

255        The authors express their sincere thanks to three anonymous reviewers for their constructive reviewing comments which helped to improve the earlier version of the manuscript. The authors also thank K. Kawasaki (Tokyo Institute of Technology), T. Koike and T. Kaneko (Yokohama National University) for their support during the experimental work and for useful discussions. This research was supported in part by a grant from the

260      Japan Society for the Promotion of Science (Y. T. and K. K).

**Supporting Information**

Preliminary data of molecular assignments by curie-point Pyr-GC/MS analysis[30] for Titan tholin products. The analytical conditions on the retention time were after the reference.[39]

265      This material is available free of charge on the Web at http://www.jsac.or.jp/analsci/.

**References**

1. J. I. Lunine and S.K. Atreya, *Nature Geosci.*, **2008**, *1*, 159.

270  2. B. N. Khare, C. Sagan, W. Thompson, E. Arakawa, F. Suits, T. Callcott, M. Williams, S. Shrader, H. Ogino, and T. Willingham, *Adv. Space Res.*, **1984**, *4*, 59.

3. B. N. Khare, C. Sagan, B. Nagy, and H. ErKarl, *Icarus,* **1986**, *68*, 176.

4. C. Sagan, W. R. Thompson, and B. N. Khare, *Accounts Chem. Res.,* **1992**, *25*, 286.

5. M. E. Brown, A. H. Bouchez, and C. A. Griffith, *Nature,* **2002**, *420*, 795.

275  6. P. Rannou, F. Hourdin, and C. McKay, *Nature,* **2002**, *418*, 853.

7. P. Rannou, F. Montmessin, F. Hourdin, and S. Lebonnois, *Science,* **2006**, *311*, 201.

8. C. A. Griffith, P. Penteado, K. Baines, P. Drossart, J. Barnes, G. Bellucci, J. Bibring, R. Brown, B. Buratti, and F. Capaccioni, *Science,* **2005**, *310*, 474.

9. D. B. Campbell, G. J. Black, L. M. Carter, and S. J. Ostro, *Science,* **2003**, *302*, 431.

280  10. C. A. Griffith, T. Owen, T. R. Geballe, J. Rayner, and P. Rannou, *Science,* **2003**, *300*, 628.

11. J. W. Barnes, R. H. Brown, E. P. Turtle, A. S. McEwen, R. D. Lorenz, M. Janssen, E. L. Schaller, M. E. Brown, B. J. Buratti, and C. Sotin, *Science,* **2005**, *310*, 92.

12. E. Stofan, C. Elachi, J. Lunine, R. Lorenz, B. Stiles, K. Mitchell, S. Ostro, L. Soderblom, C. Wood, and H. Zebker, *Nature,* **2007**, *445*, 61.

285  13. J. I. Lunine and R. D. Lorenz, *Ann. Rev. Earth Planet. Sci.,* **2009**, *37*, 299.

14. C. C. Porco, E. Baker, J. Barbara, K. Beurle, A. Brahic, J. A. Burns, S. Charnoz, N. Cooper, D. D. Dawson, and A. D. Genio, *Nature,* **2005**, *434*, 159.

15. J. P. Lebreton, O. Witasse, C. Sollazzo, T. Blancquaert, P. Couzin, A. M. Schipper, J. B. Jones, D. L. Matson, L. I. Gurvits, and D. H. Atkinson, *Nature,* **2005**, *438*, 758.

**Figure Captions**

Fig. 1 **(a)** Profile showing the pressure, temperature, and energy sources available in Titan's atmosphere (*i.e.*, Saturnian magnetospheric electrons, long ultraviolet light (>155 nm), and cosmic ray[s]).[1,35,36] The altitudes of maximum energy deposition are shown by "+" symbols.[26] **(b)** Two-dimensional planetary distribution of haze scaled extinction. The winds move from the south polar (summer) towards the north polar region (winter).[6]

Fig. 2 Photographs of the apparatus used at the Tokyo Institute of Technology to produce 3 MeV proton irradiation in order to simulate Titan's atmosphere. **(a)** Two seconds after initiation of proton irradiation. **(b)** Ten seconds after proton irradiation, a white–yellow misty haze has formed. Please see also, energy yields normalized by amino acids[37-39] and schematic diagram of the experimental setup for proton irradiation using a van de Graaff accelerator.[40-41]

Fig. 3 **(a)** Representative scanning electron microscopy (SEM) images of aggregated organic complexes (*i.e.*, simulated Titan tholins) formed from the mixture of nitrogen and methane by proton irradiation. **(b)** Magnified view of the area outlined by the dashed rectangle in Fig. 3a. **(c, d, e)** Other SEM images of the complex organic aggregates present in the simulated Titan tholins. The SEM imaging samples (**a** - **e**) were recovered by water to simulate the morphology of Tholin after eventual water exposure (cf. suggestion from Griffith et al.[10]).

Fig. 4 Three-dimensional atomic force microscopy (AFM) images of the complex organic aggregates in the simulated Titan tholins that were formed from the mixture of nitrogen and methane by proton irradiation. The AFM system used was a SII SPA 400 AFM unit (Seiko Instruments). Three dimensional (3D) images of Figs. 4a and 4c were measured for morphological size (nm) on Figs. 4b and 4d, respectively. The measurement positions (colored in red, yellow, blue, green, purple) in the scale of nanometer (nm) were assigned by Z1, Z2, height, and its distance. The AFM imaging samples were recovered by water to simulate the morphology of Tholin after eventual water exposure (cf. suggestion from Griffith et al.[10]).

Fig. 5 Gas chromatograph of Titan tholins after pyrolysis at 600°C. The inset shows the average mass spectrum of the main peak ($m/z$ = 41 and 44 were unidentified). The scanning region ranged from $m/z$ = 10 to 100. Other spectroscopic analyses[52] were used for identification of $NH_3$ and HCN. See also, preliminary data of molecular assignments by Curie-point Pyr-GC/MS analysis[30] for Titan tholin product in the Supplementary Information. The analytical condition on the retention time was after the reference.[39]

Fig. 6 **(a)** Gas chromatographic scan of the simulated Titan tholins after acid hydrolysis. **(b)** Mass spectra of standard glycine[44] and synthesized glycine formed from a mixture of nitrogen and $^{13}C$-labeled methane by proton irradiation. **(c)** The chemical structures of the detected amino acids.

Fig. 7 Matrix-assisted laser desorption ionization–time of flight–mass spectrometry traces of glycine (Gly, [M+H]$^+$) and alanine (Ala, [M+H]$^+$) produced from the simulated Titan tholins. **(a)** The front spectra show the *m/z* for glycine and alanine hydrolyzed in the presence of ordinary H$_2^{16}$O. **(b)** The back spectra show the *m/z* for glycine and alanine hydrolyzed with $^{18}$O-labeled H$_2^{18}$O. Due to the ionization efficiency of the matrix-assisted laser desorption ionization technique, we verified the total molecular weight of tholin by GPC analysis.

Fig. 8 Gel permeation chromatogram (GPC) of Titan tholins formed by proton irradiation. The dashed line stands for the calibration curve corresponding to the molecular weight (right-hand axis). GPC area-based relative abundances were as follows: fraction A (0.2%); B (0.2%); C (79.1%); D (16.7%); E (3.8%).

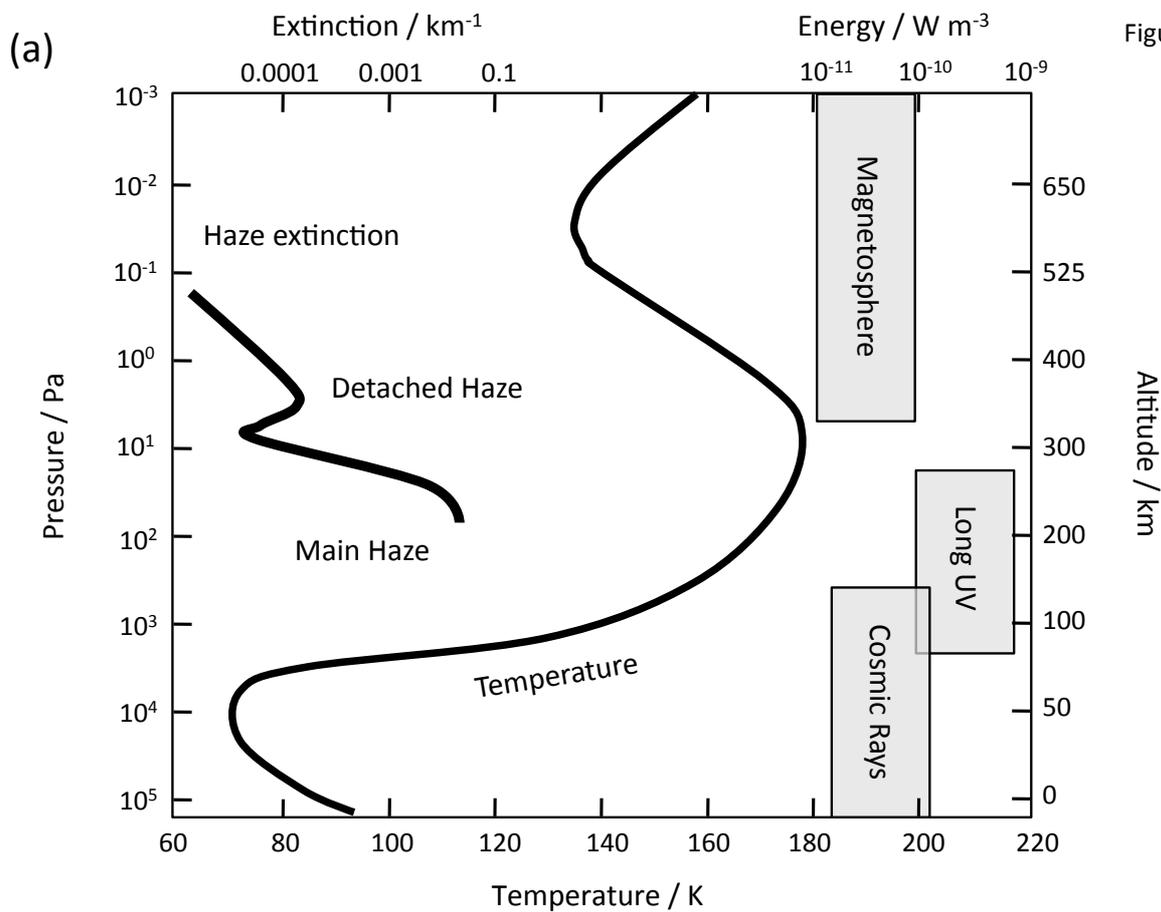

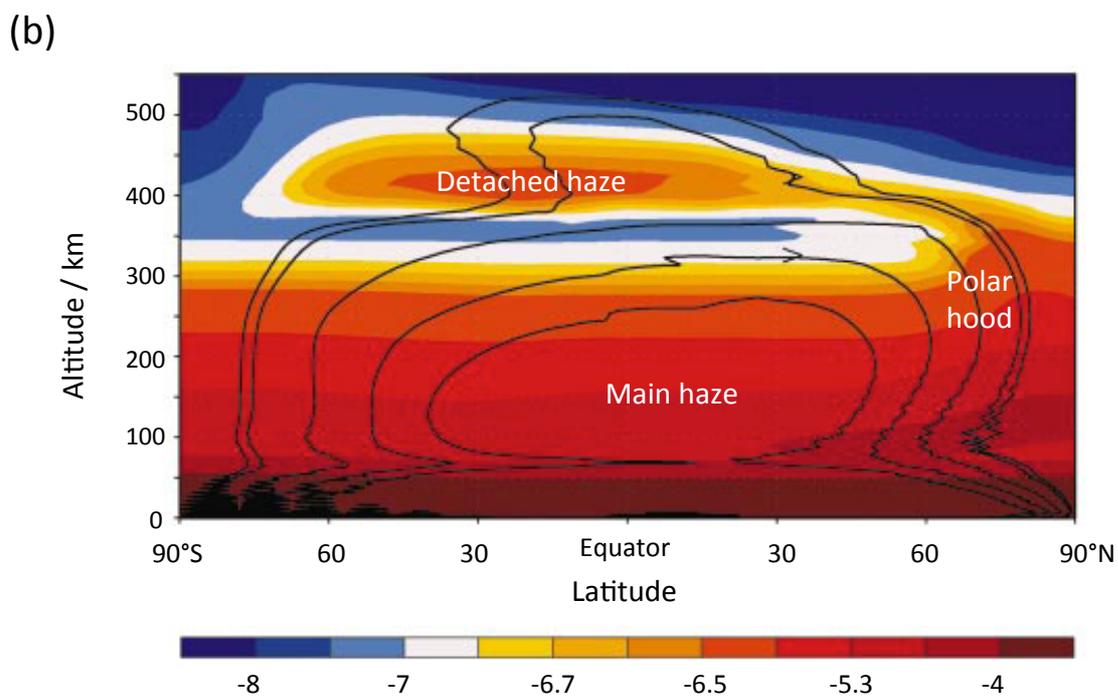

Figure 1

Figure 2

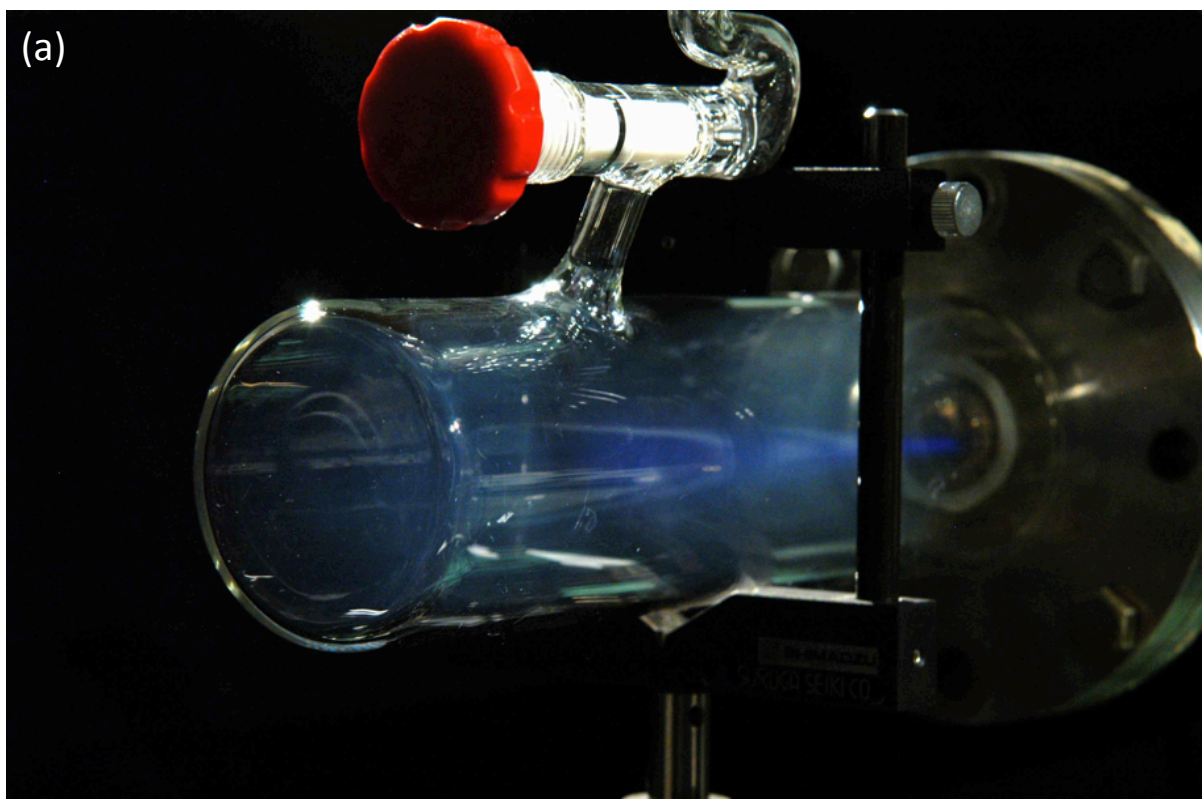

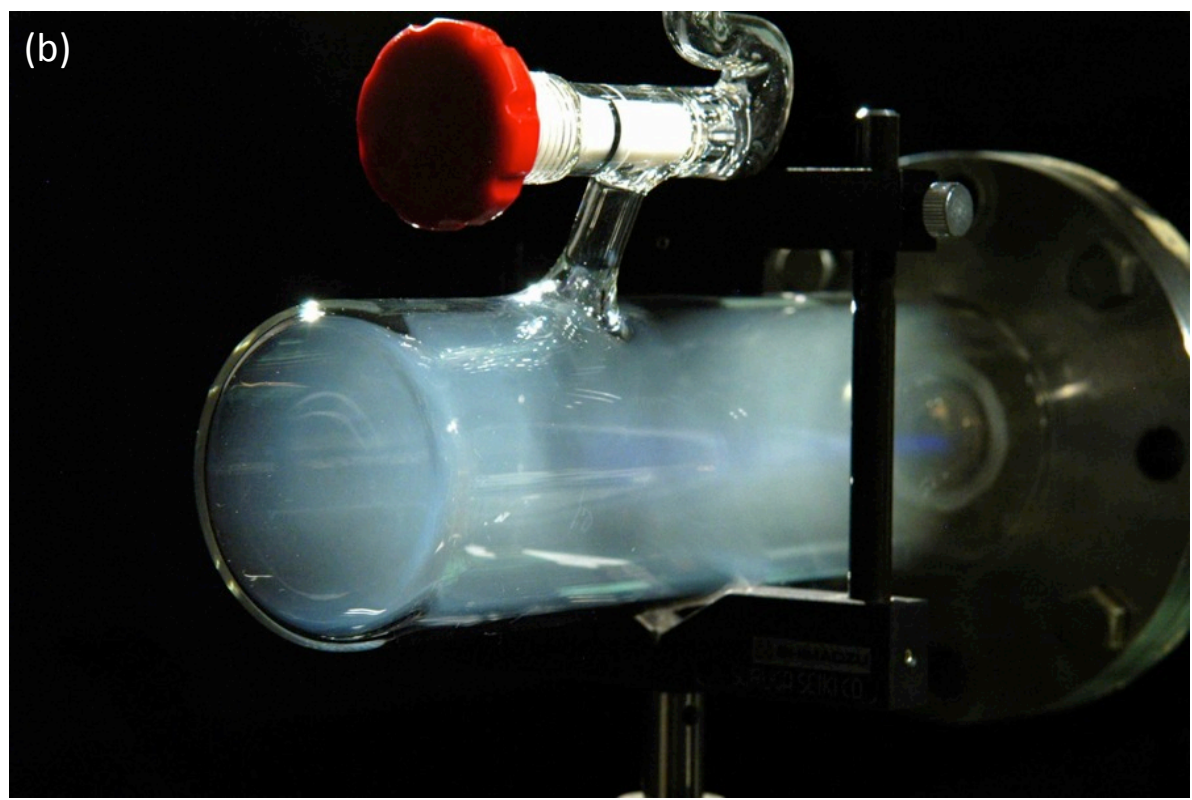

Figure 3

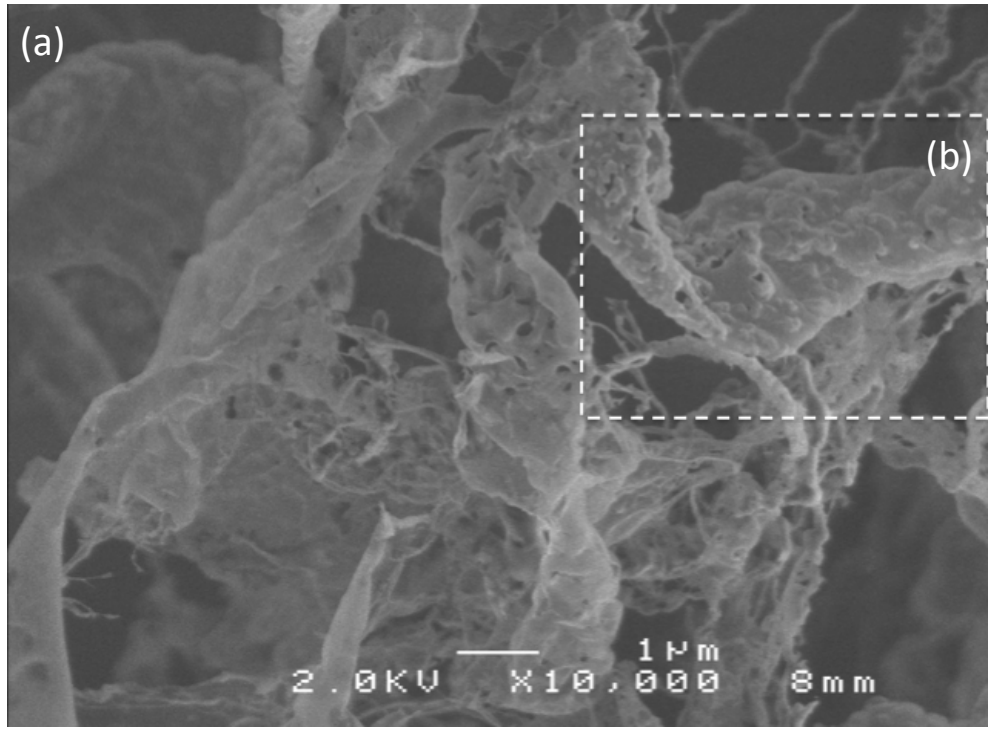

(a)

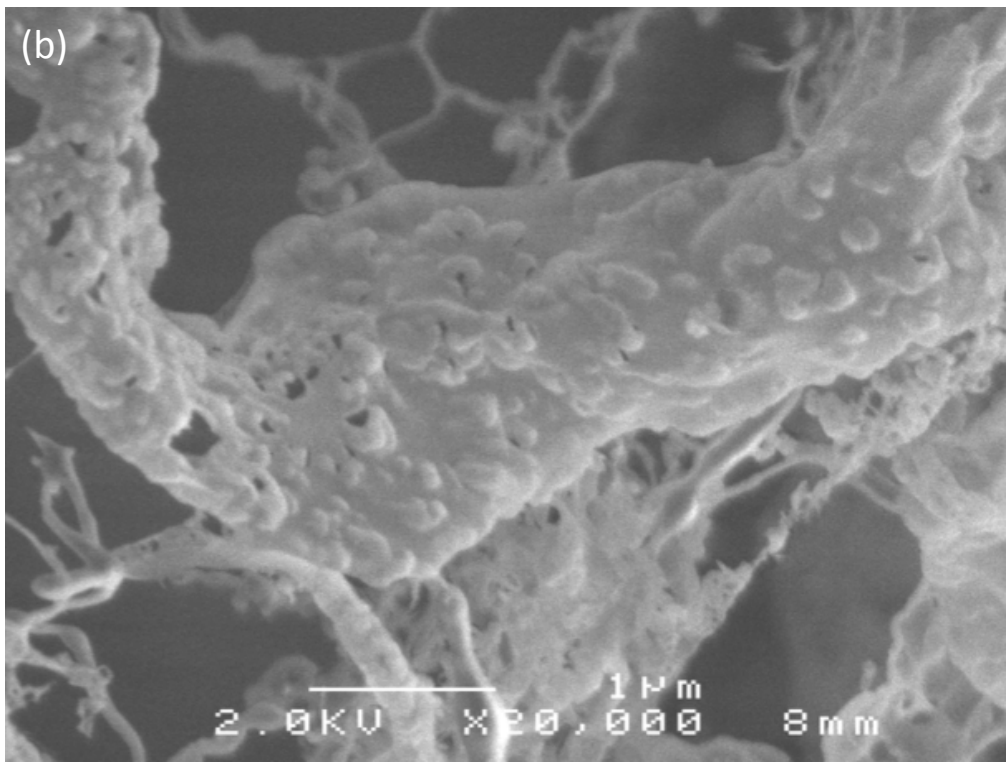

(b)

Figure 3

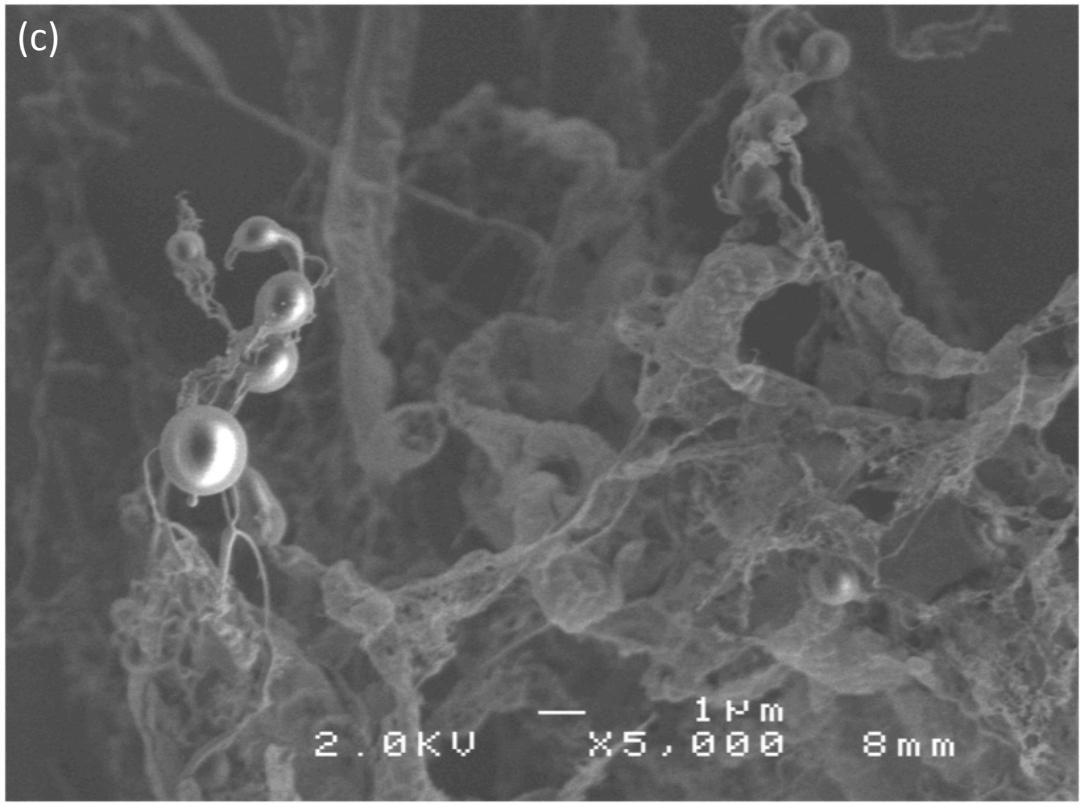

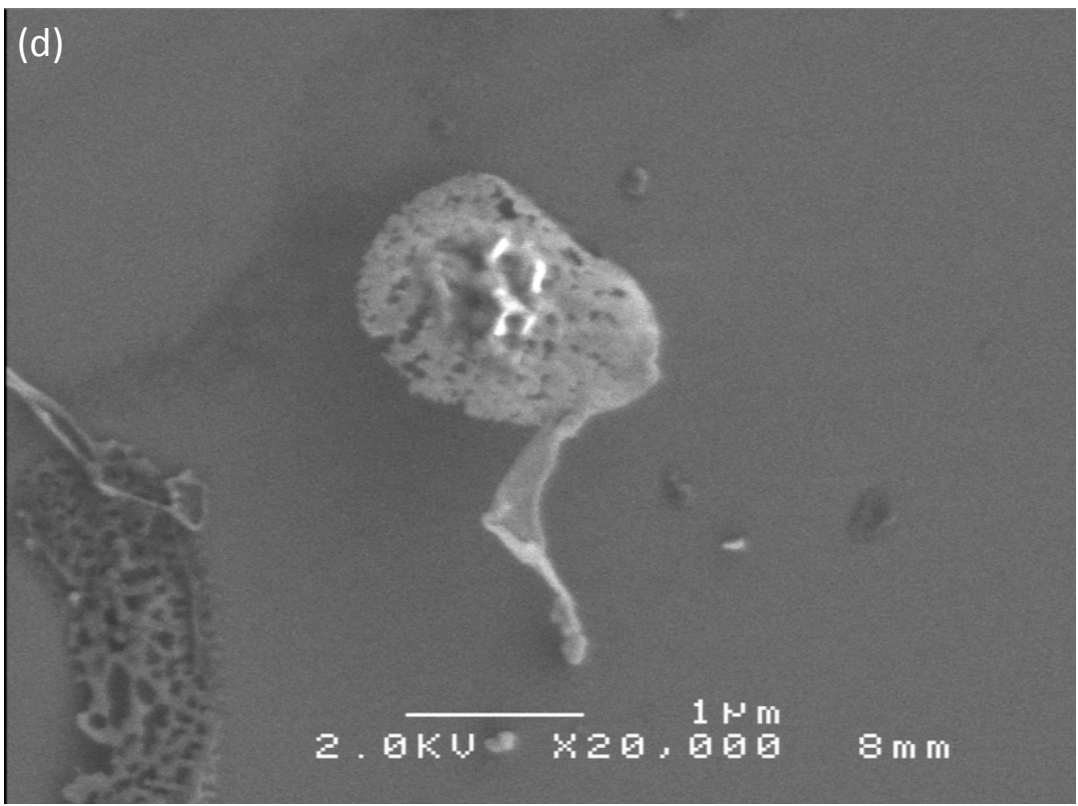

Figure 3

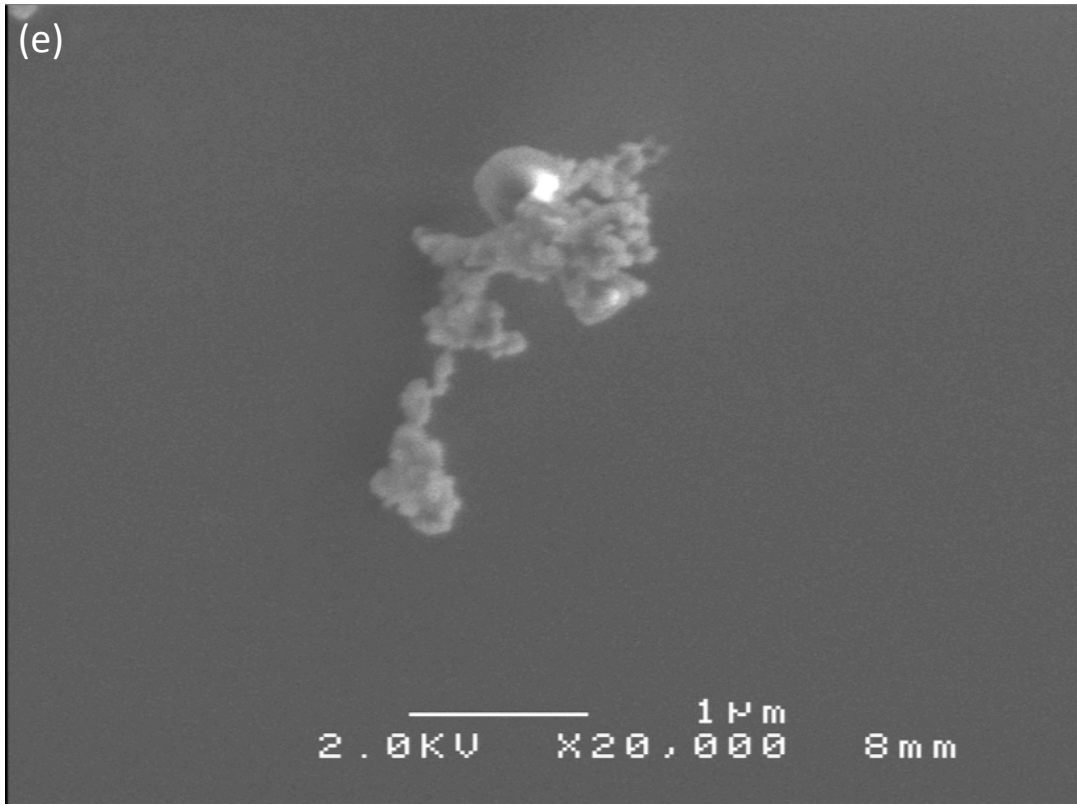

Figure 4

(a)

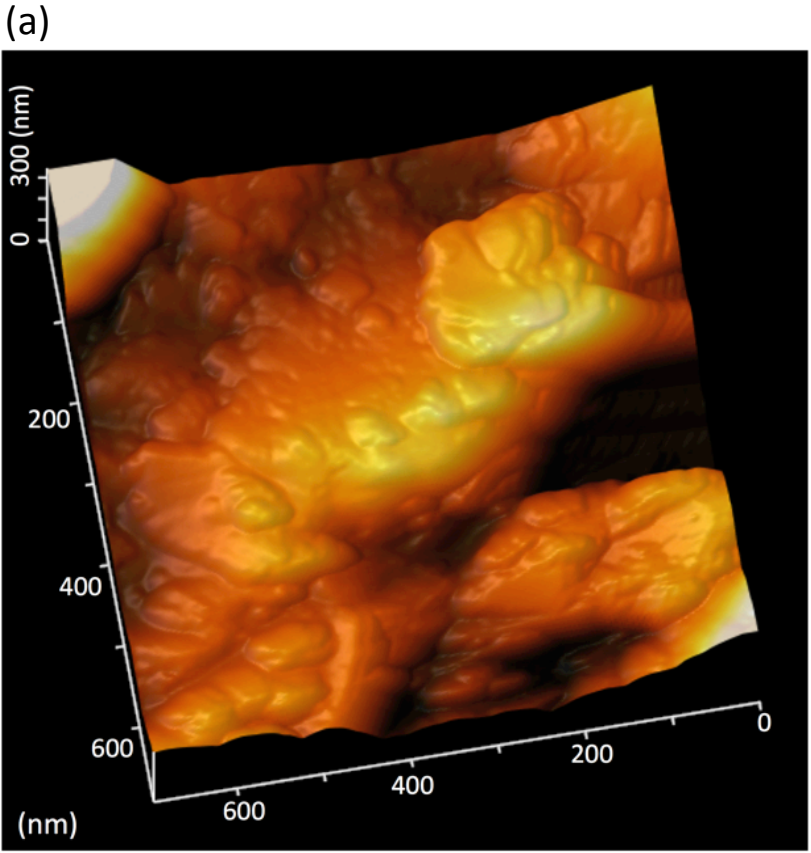

(b)

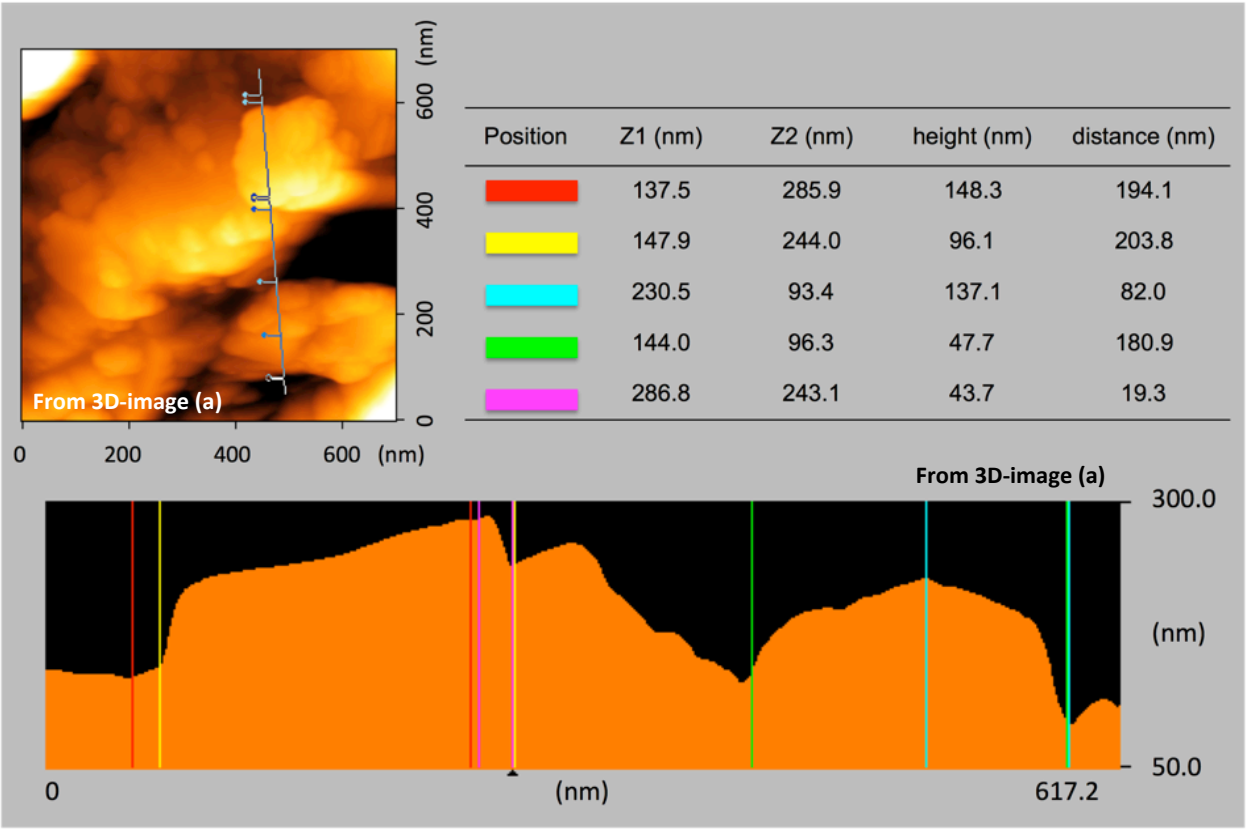

Figure 4

(c)

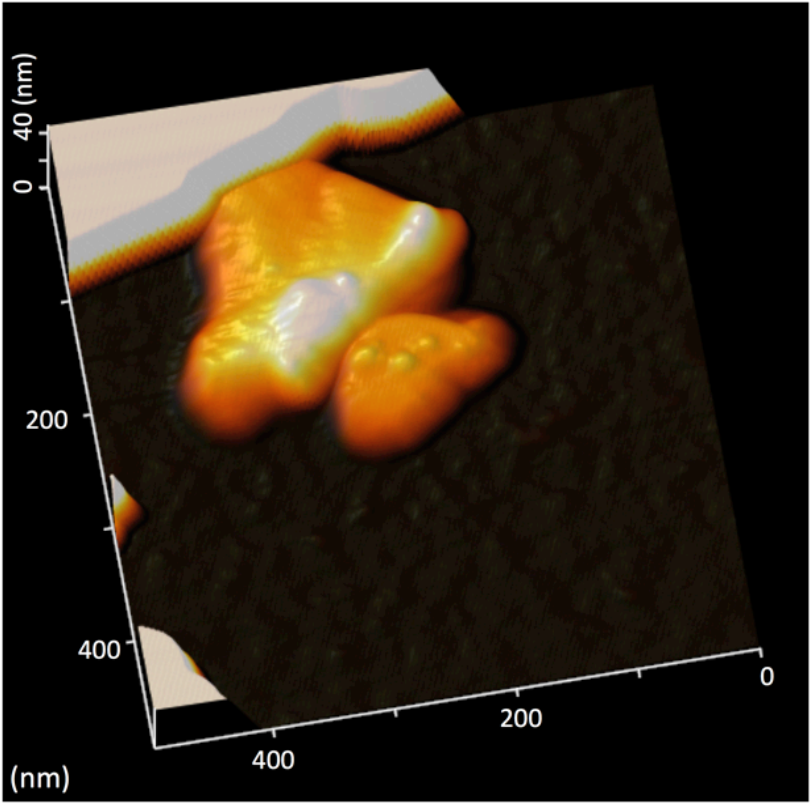

(d)

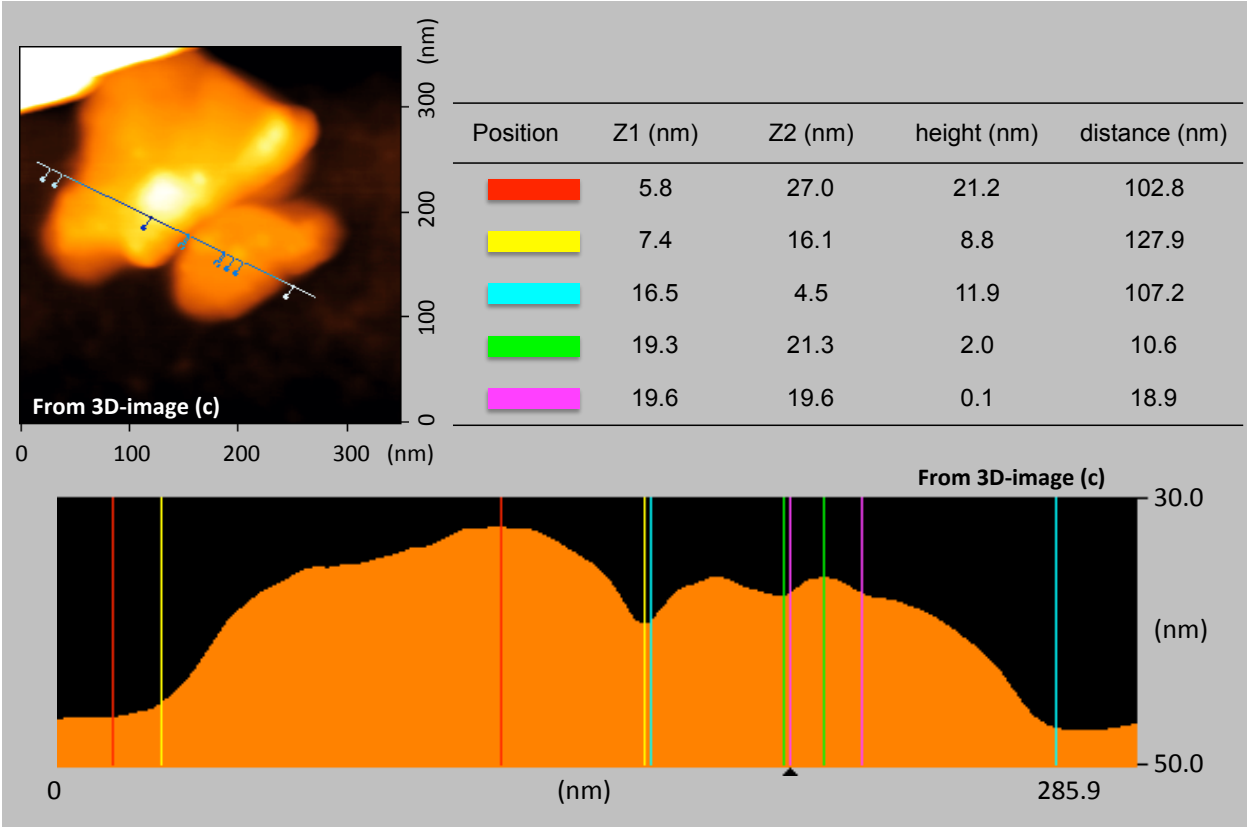

| Position | Z1 (nm) | Z2 (nm) | height (nm) | distance (nm) |
|---|---|---|---|---|
| 🟥 | 5.8 | 27.0 | 21.2 | 102.8 |
| 🟨 | 7.4 | 16.1 | 8.8 | 127.9 |
| 🟦 | 16.5 | 4.5 | 11.9 | 107.2 |
| 🟩 | 19.3 | 21.3 | 2.0 | 10.6 |
| 🟪 | 19.6 | 19.6 | 0.1 | 18.9 |

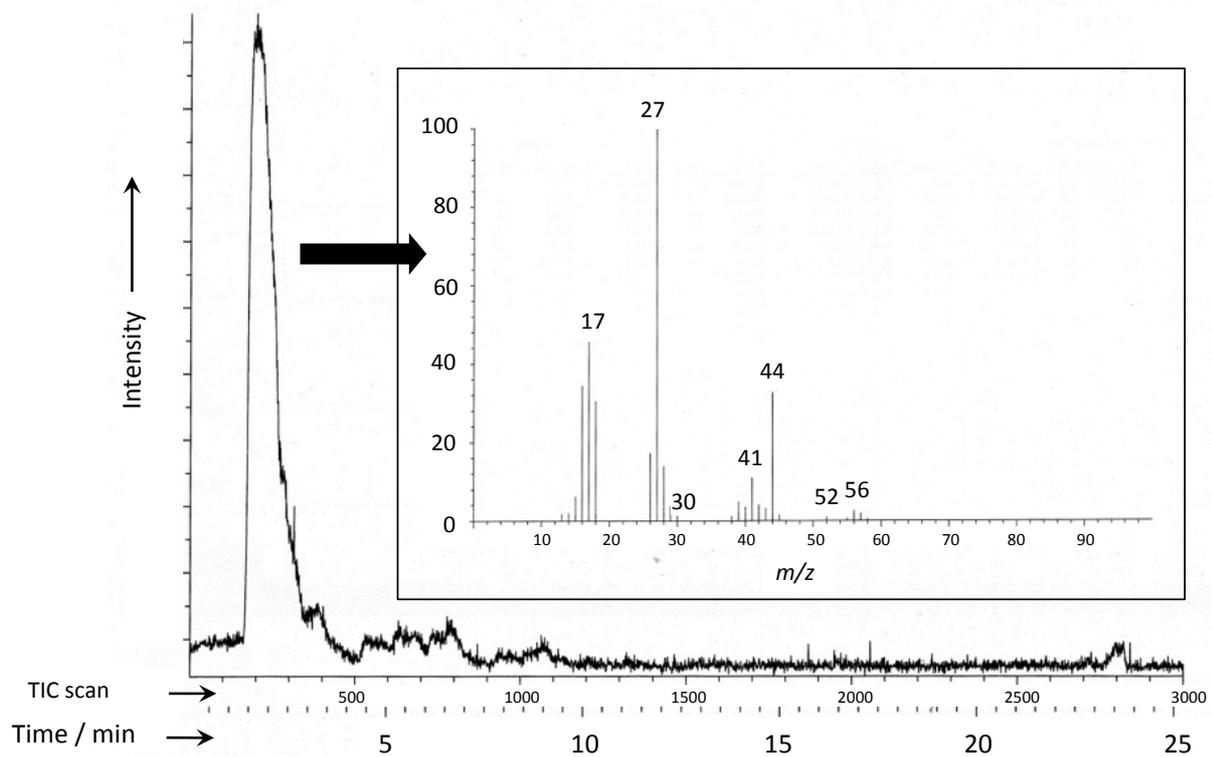

Figure 5

Figure 6

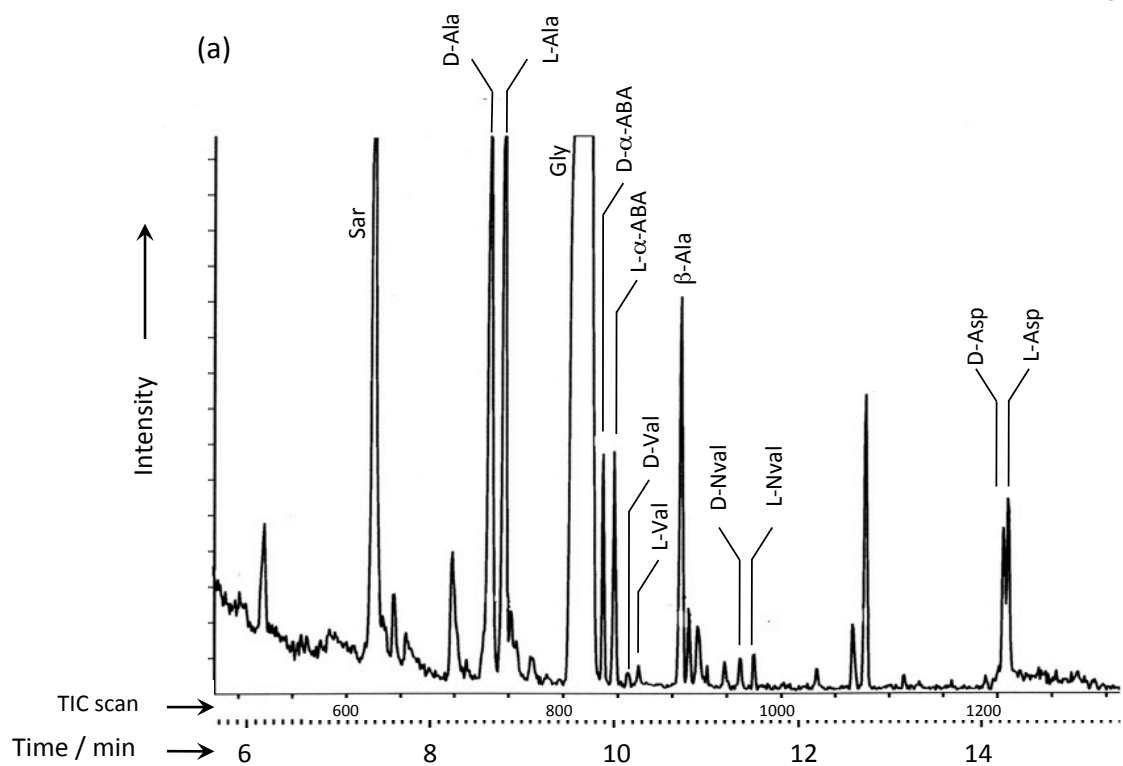

(a)

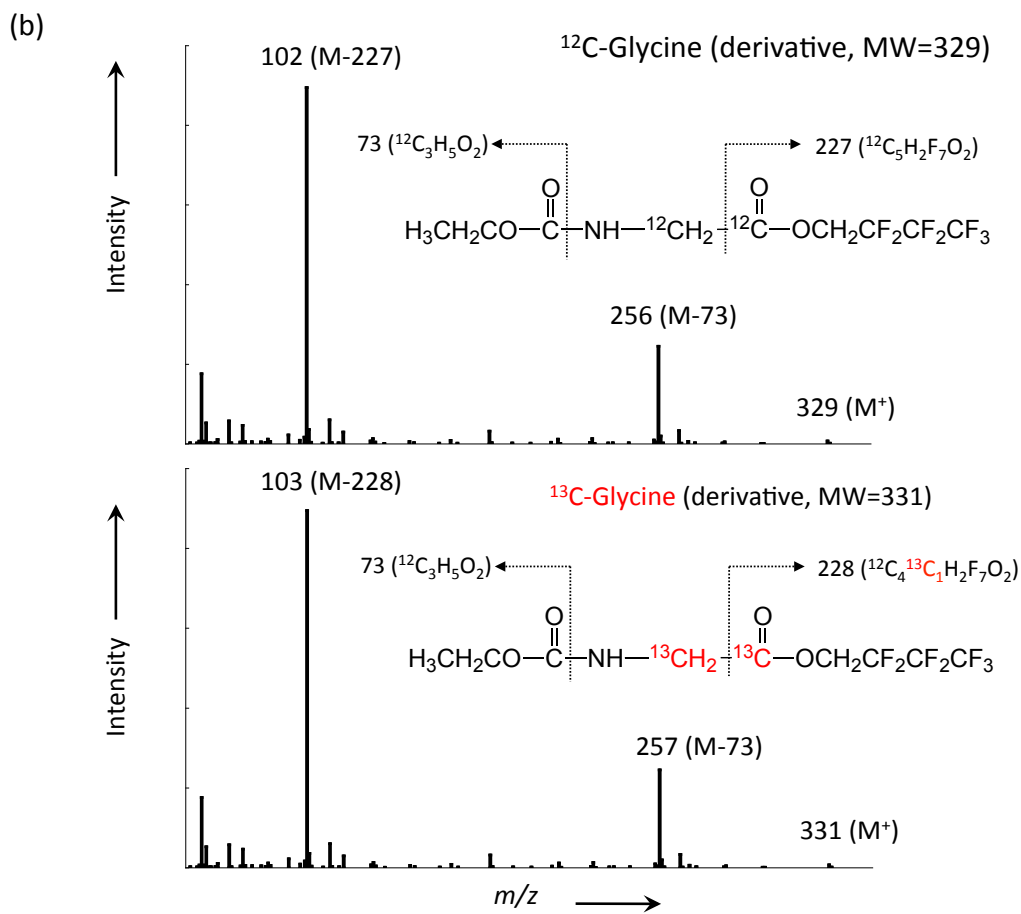

(b)

(c)  Figure 6

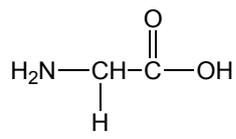 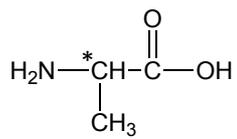 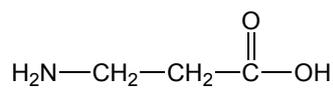 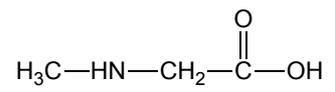

Glycine      α-D-,L-Alanine      β-Alanine      Sarcosine
(Gly: $C_2$)      (α-Ala: $C_3$)      (β-Ala: $C_3$)      (Sar: $C_3$)

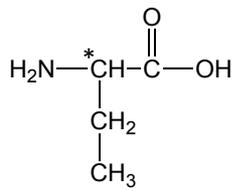 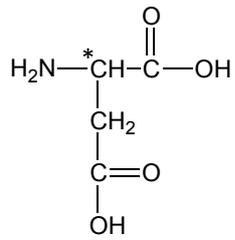 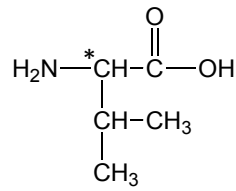 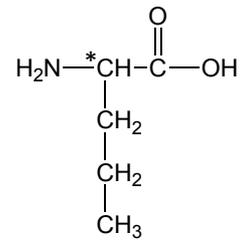

α-D-,L-Aminobutylic acid      D-,L-Aspartic acid      D-,L-Valine      D-,L-Norvaline
(α-ABA: $C_4$)      (Asp: $C_4$)      (Val: $C_5$)      (Nval: $C_5$)

\* chiral center

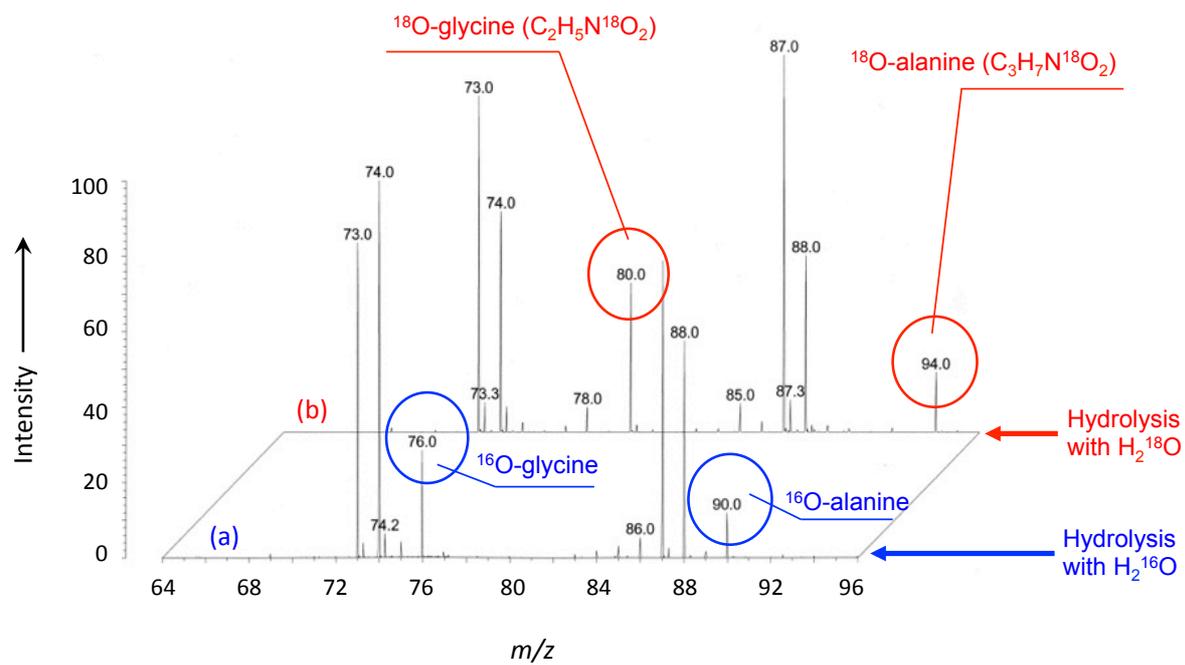

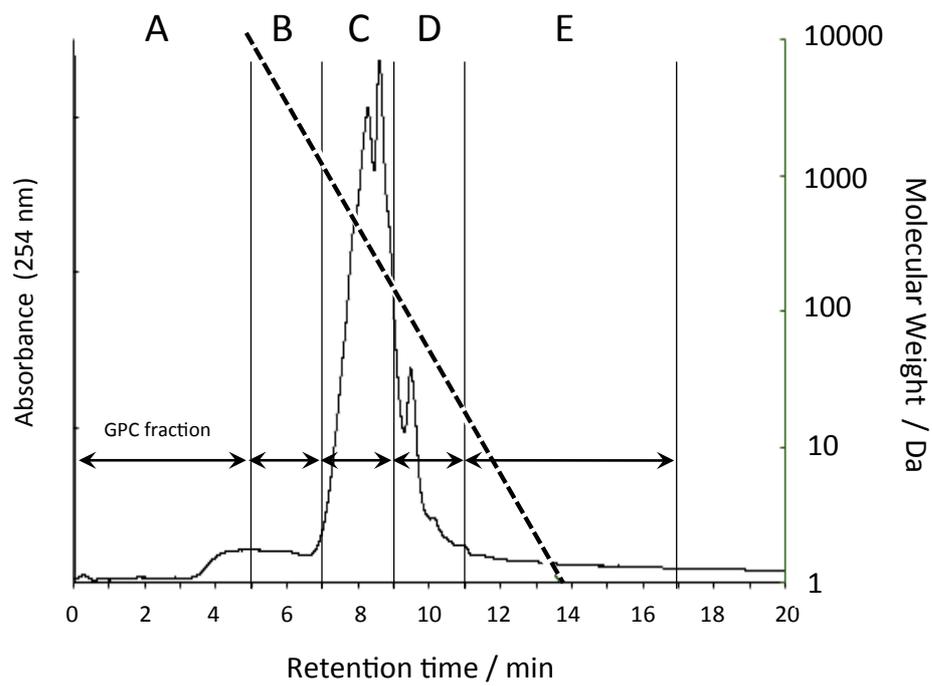

Figure 8